\documentstyle[12pt]{article}
\input psfig

\renewcommand{\baselinestretch}{1.2}
\newcommand{\nc}{\newcommand}

\nc{\eqr}[1]{(\ref{#1})}
\nc{\sref}[1]{\S\ref{#1}}
\nc{\tref}[1]{Table~\ref{#1}}
\nc{\fref}[1]{Figure~\ref{#1}}
\nc{\cref}[1]{Chapter~\ref{#1}}
\nc{\beq}{\begin{equation}}
\nc{\eeq}{\end{equation}}
\nc{\barray}{\begin{eqnarray}}
\nc{\earray}{\end{eqnarray}}
\nc{\barrayn}{\begin{eqnarray*}}
\nc{\earrayn}{\end{eqnarray*}}
\nc{\bcenter}{\begin{center}}
\nc{\ecenter}{\end{center}}
\nc{\lra}{\longrightarrow}
\nc{\ra}{\rightarrow}
\nc{\setall}{\setcounter{equation}{0}
        \setcounter{definition}{0}
        \setcounter{lemma}{0}
        \setcounter{convention}{0}
        \setcounter{conjecture}{0}
        \setcounter{theorem}{0}
        \setcounter{proposition}{0}
        \setcounter{property}{0}
        \setcounter{fact}{0}
        \setcounter{corollary}{0}}
\nc{\setequation}{\setcounter{equation}{0}}
\nc{\hs}[1]{\hspace{#1 mm}}
\newcommand{\rtimes}{\mbox{$\times\!\rule{0.3pt}{1.1ex}\,$}}

\def\sCC{{\kern 0.27em\vrule height1.45ex width0.03em depth0em
          \kern-0.30em\rm C}}
\def\C{{\mathchoice
  {\sCC}
  {\sCC}
  {\kern 0.225em \vrule height1.05ex width0.025em depth0em \kern-0.25em \rm C}
  {\kern 0.180em \vrule height0.78ex width0.02em depth0em \kern-0.2em \rm C}
        }}
\def\sHH{{\rm I\kern-.16em{}H}}
\def\H{{\mathchoice
  {\sHH}
  {\sHH}
  {\rm I\kern-.13em{}H}
  {\rm I\kern-.13em{}H} }}
\def\sNN{{\rm I\kern-.16em{}N}}
\def\N{{\mathchoice
  {\sNN}
  {\sNN}
  {\rm I\kern-.12em{}N}
  {\rm I\kern-.10em{}N} }}
\def\sPP{{\rm I\kern-.16em{}P}}
\def\P{{\mathchoice
  {\sPP}
  {\sPP}
  {\rm I\kern-.12em{}P}
  {\rm I\kern-.10em{}P} }}
\def\sQQ{{\kern 0.27em \vrule height1.45ex width0.03em depth0em
          \kern-0.30em \rm Q}}
\def\Q{{\mathchoice
        {\sQQ}
        {\sQQ}
  {\kern 0.225em \vrule height1.05ex width0.025em depth0em \kern-0.25em \rm Q}
  {\kern 0.180em \vrule height0.78ex width0.020em depth0em \kern-0.20em \rm Q}
        }}
\def\sRR{{\rm I\kern-0.16em{}R}}
\def\R{{\mathchoice
  {\sRR}
  {\sRR}
  {\rm I\kern-0.12em{}R}
  {\rm I\kern-0.10em{}R} }}
\def\sZZ{{\rm Z\kern-0.32em{}Z}}
\def\Z{{\mathchoice
  {\sZZ}
  {\sZZ} 
  {\rm Z\kern-0.3em{}Z}     
  {\rm Z\kern-0.25em{}Z} }}  
\def\ZZZ{{\rm Z\kern-0.24em{}Z}}
\def\sKK{{\rm I\kern-0.16em{}K}}
\def\K{{\mathchoice
  {\sKK}
  {\sKK}
  {\rm I\kern-0.12em{}K}
  {\rm I\kern-0.10em{}K} }}
\def\odd{\rm{odd}}
\def\even{\rm{even}}
\def\G{$Z_k\times D_{k'}~$}

\newtheorem{lemma}{\bf LEMMA}

\newtheorem{proposition}{\bf PROPOSITION}

\begin{document}
\renewcommand{\baselinestretch}{1}

\thispagestyle{empty}
{\flushright{\small MIT-CTP-2872\\hep-th/9906031\\}}

\vspace{.3in}
\begin{center}\LARGE {The $Z_k \times D_{k'}$ Brane Box Model}
\end{center}

\vspace{.2in}
\begin{center}
{\large Bo Feng, Amihay Hanany and Yang-Hui He\\}
\normalsize{fengb, hanany, yhe@ctp.mit.edu\footnote{
Research supported in part
by the CTP and the LNS of MIT and the U.S. Department of Energy 
under cooperative research agreement \# DE-FC02-94ER40818; YHH is also
supported by the NSF Graduate Fellowship.}
\\}
\vspace{.2in} {\it Center for Theoretical Physics,\\ Massachusetts
Institute of Technology\\ Cambridge, Massachusetts 02139, U.S.A.\\}
\end{center}
\vspace{0.1in}

\begin{abstract}

An example of a non-Abelian Brane Box Model, namely 
one corresponding to a $Z_k \times D_{k'}$ orbifold singularity of
$\C^3$, is constructed. Its self-consistency and hence equivalence to
geometrical methods are subsequently shown. 
It is demonstrated how a group-theoretic twist of the
non-Abelian group circumvents the problem of inconsistency that arise
from na\"{\i}ve attempts at the construction.
\end{abstract}

\section{Introduction}
Brane setups \cite{Han-Wit} have been widely attempted to 
provide an alternative to algebro-geometric methods in the construction of gauge 
theories (see \cite{Giveon} and references therein). The advantages of the latter 
include the enlightening of important properties of manifolds such as mirror
symmetry, the provision of convenient supergravity descriptions and in instances
of pure geometrical engineering, the absence of non-perturbative objects.
The former on the other hand, give
intuitive and direct treatments of the gauge theory. One can conveniently
read out much information concerning the
gauge theory from the brane setups, such as the dimension of the Coulomb and Higgs 
branches \cite{Han-Wit}, the mirror symmetry \cite{Han-Wit,Boer,Kapustin,
P-Zaf} in 3 
dimensions first shown in \cite{IS}, 
the Seiberg-duality in 4 
dimensions \cite{Elitzur}, and exact solutions
when we lift the setups from Type IIA to M Theory \cite{Mlift}.

In particular, when discussing ${\cal N}=2$ supersymmetric gauge theories in 
4 dimensions, there are three known methods currently in favour.
The first method is {\it geometrical engineering}
exemplified by works in \cite{Mirror}; 
the second uses D3 branes as {\it probes} on orbifold singularities of 
the type 
$\C^2/\Gamma$ with $\Gamma$ being a finite discrete subgroup of $SU(2)$
\cite{Quiver}, and the third, the usage of {\it brane setups}.
These three approaches are related to each other by
proper T or S Dualities \cite{Karch}.
For example,
the configuration of stretching Type IIA D4 branes between $n+1$ NS5 
branes placed in a 
circular fashion, the so-called {\bf elliptic model}\footnote{We call it elliptic 
even though
there is only an $S^1$ upon which we place the D4 branes; this is because from the
M Theory perspective, there is another direction: an $S^1$ on which we compactify to 
obtain type Type IIA. The presence of two $S^1$'s makes the theory toroidal, or
elliptic.
Later we shall see how to make use of $T^2=S^1 \times S^1$ in Type IIB.
For clarity we shall refer to the former as the ${\cal N}=2$ elliptic model and the latter,
the ${\cal N}=1$ elliptic model.}, 
is precisely T-dual to D3 branes stacked upon ALE\footnote{ 
Asymptotically Locally Euclidean, i.e., Gorenstein singularities that locally represent
Calabi-Yau manifolds.} singularities of type $\widehat{A_n}$ (see \cite{Mlift,
Karl,B-Karch,Park,Erlich} for detailed discussions).

The above constructions can be easily generalised to ${\cal N}=1$ 
supersymmetric field theories in 4 dimensions. 
Methods from geometric engineering as well as D3 branes as probes
now dictate the usage of orbifold singularities of the type 
$\C^3/\Gamma$ with $\Gamma$ being a finite discrete subgroup of 
$SU(3)$ \cite{Conf,Quiver1}.
A catalogue of all the discrete subgroups of $SU(3)$ in this context
is given in \cite{Han-He,Muto1}.
Now from the brane-setup point of view, there are two ways to arrive at 
the theory. The first is to rotate certain branes in the configuration to
break the supersymmetry from ${\cal N}=2$ to ${\cal N}=1$ \cite{Elitzur}. 
The alternative is to add
another type of NS5 branes, viz., a set of NS5$'$ branes placed
perpendicularly to the original NS5, whereby
constructing the so-called {\bf Brane Box Model} \cite{Han-Zaf,Han-Ura}. 
Each of these two different approaches has its own merits. 
While the former (rotating branes) facilitates the deduction of Seiberg Duality, 
for the latter (Brane Box Models), it is easier to construct a class of new,
finite, chiral 
field theories \cite{Han-S}.  By finite we mean that in the field theory the 
divergences may be cancelable.
From the perspective of branes on geometrical singularities, 
this finiteness corresponds to the
cancelation of tadpoles in the orbifold background and from that of
brane setups, it corresponds to the no-bending requirement of 
the branes \cite{Karch,Han-S,Leigh}. 
Indeed, as with the ${\cal N}=2$ case, we can still show 
the equivalence among these
different perspectives by suitable S or T Duality transformations. 
This equivalence is explicitly shown in \cite{Han-Ura} for the case of 
the Abelian finite subgroups of $SU(3)$.
More precisely, for the group $Z_k \times Z_{k'}$ or $Z_k$ 
and a chosen decomposition
of ${\bf 3}$ into appropriate irreducible representations thereof
one can construct 
the corresponding Brane Box Model that gives the
same quiver diagram as the one 
obtained directly from the geometrical methods of
attack; this is what we mean by equivalence \cite{Conf}. 

Indeed, we are not satisfied with the fact that this abovementioned equivalence
so far exists only for Abelian singularities and would like to see how it may be
extended to non-Abelian cases.
The aim for constructing Brane Box Models of non-Abelian finite 
groups is twofold: firstly we would generate a new category of finite supersymmetric
field theories and secondly we would demonstrate how the equivalence between the 
Brane Box Model and D3 branes as probes is true beyond the Abelian case and hence
give an interesting physical perspective on non-Abelian groups.
More specifically, the problem we wish to tackle is that given any finite
discrete subgroup $\Gamma$ of $SU(2)$ or $SU(3)$,
what is the brane setup (in the T-dual picture)
that corresponds to D3 branes as probes on orbifold singularities afforded by
$\Gamma$?
For the $SU(2)$ case, the answer for the $\widehat{A}$ series
was given in \cite{Mlift} and that for the $\widehat{D}$ series,
in \cite{Kapustin}, yet $\widehat{E_{6,7,8}}$ are still unsolved.
For the $SU(3)$ case, the situation is even worse.
While \cite{Han-Zaf,Han-Ura} have given solutions to the Abelian groups
$Z_k$ and $Z_k\times Z_{k'}$, the non-Abelian $\Delta$ and $\Sigma$ series
have yet to be treated.
Though it is not clear how the generalisation can be done for
arbitrary non-Abelian singularities,
it is the purpose of this writing to take one further step from \cite{Han-Zaf,Han-Ura},
and address the next simplest series of dimension three
orbifold theories, viz., those of $\C^3/Z_k \times D_{k'}$ and
construct the corresponding
Brane Box Model and show its equivalence to geometrical methods. In addition to
equivalence we demonstrate how the two pictures are bijectively related for the
group of interest and that given one there exists a unique description in the other.
The key input is given by Kutasov, Sen and Kapustin in \cite{Kapustin,Kutasov,Sen}.
Moreover \cite{Han-Zaf2} 
has briefly pointed out how his results may be used, but
without showing the consistency and equivalence.

The paper is organised as follows.
In section \sref{sec:review} we shall briefly review some techniques of brane setups
and orbifold projections in the context of finite quiver theories. Section
\sref{sec:group} is then devoted to a crucial digression on the mathematical properties
of the group of our interest, or what we call $G:=Z_k \times D_{k'}$. In section
\sref{sec:BB} we construct the Brane Box Model for $G$, followed by concluding
remarks in section \sref{sec:conc}.

\section*{Nomenclature}
Unless otherwise stated, we shall, throughout our paper, adhere to the notation
that $\omega_n = e^{\frac{2 \pi i}{n}}$, the $n$th root of unity,
that $G$ refers to the group \G, that without ambiguity
$Z_k$ denotes $\Z_k$, the cyclic group of $k$ elements, that $D_k$ is
the binary dihedral group of order $4k$ and gives the affine Dynkin
diagram of $\widehat{D}_{k+2}$,
and that $d_k$ denotes
the ordinary dihedral group of order $2k$. Moreover $\delta$ will be defined as
$(k,2k')$, the greatest common divisor (GCD) of $k$ and $2k'$. 

\section{A Brief Review of $D_n$ Quivers, Brane Boxes, 
	and Brane Probes on Orbifolds} \label{sec:review}
The aim of our paper is to construct the Brane Box Model of the 
non-Abelian finite group \G and to show its consistency as well as
equivalence to geometric methods. To do so,
we need to know how to read out the gauge groups and matter content
from quiver diagrams which describe a particular 
field theory from the geometry side. 
The knowledge for such a task is supplied in \sref{subsec:Quiver}.
Next, as mentioned in the introduction, to construct field theories
which could be encoded in the $D_k$ quiver diagram, 
we need an important result from \cite{Kapustin,Kutasov,Sen}.
A brief review befitting
our aim is given in \sref{subsec:Kapustin}. 
Finally in \sref{subsec:BBZZ} we present the rudiments of the
Brane Box Model.

\subsection{Branes on Orbifolds and Quiver Diagrams} \label{subsec:Quiver}
It is well-known that a stack of coincident $n$ D3 branes gives rise to an ${\cal N}=4$ 
$U(n)$ super-Yang-Mills theory on the four dimensional world volume. The $U(1)$ factor
of the $U(n)$ gauge group decouples when we discuss the low energy dynamics of 
the field theory and can be ignored, therefore giving us an effective $SU(n)$ theory.
For ${\cal N}=4$ in 4 dimensions the R-symmetry is $SU(4)$. Under such an R-symmetry, 
the fermions in the vector multiplet transform in
the spinor representation of $SU(4) \simeq Spin(6)$ and the scalars,
in the vector representation of $Spin(6)$, the universal cover of $SO(6)$. 
In the brane picture we can identify the R-symmetry as the $SO(6)$ 
isometry group which acts on the six transverse directions of the D3-branes.
Furthermore, in the AdS/CFT picture, 
this $SU(4)$ simply manifests as the $SO(6)$ isometry group of the
5-sphere in $AdS_{5}\times S^{5}$ \cite{Conf}.

We shall refer to this gauge theory of the D3 branes as the parent theory and
consider the consequences of putting the stack on geometric singularities.
A wide class of finite Yang-Mills theories of various
gauge groups and supersymmetries is obtained when the parent theory is placed on orbifold
singularities of the type $\C^m/\Gamma$ where $m=2,3$.
What this means is that we select a discrete finite group $\Gamma \subset SU(4)$
and let its irreducible representations $\{{\bf r}_i\}$ act on the Chan-Paton 
indices $I,J=1,...,n$ 
of the D3 branes by permutation. Only those matter fields of the parent theory
that are invariant under the group action of $\Gamma$ remain, the rest are
eliminated by this so-called ``orbifold projection''.
We present the properties of the parent and the orbifolded theory in the following
diagram:
\[
\begin{array}{|l|lll|}
\hline
&$Parent Theory$	& \stackrel{\Gamma,{\rm~irreps~}=\{{\bf r}_i\}}{\longrightarrow}
			&$Orbifold Theory$\\
\hline
$SUSY$			&{\cal N}=4	&	&
	\begin{array}{l}
	{\cal N}=2, {\rm~for~} \C^2/\{\Gamma\subset SU(2)\}	\\
	{\cal N}=1, {\rm~for~} \C^3/\{\Gamma\subset SU(3)\}	\\
	{\cal N}=0, {\rm~for~} (\C^3\simeq\R^6)/\{\Gamma\subset \{SU(4)\simeq SO(6)\}\}	\\
	\end{array}\\
\hline
\begin{array}{c}
	$Gauge$ \\
	$Group$
\end{array}		&U(n)		&	& \prod\limits_{i} SU(N_i), 
		{\rm~~~~~~~where~}\sum\limits_{i} N_i \dim{\bf r}_i = n\\
\hline
$Fermion$	&\Psi_{IJ}^{\bf{4}} &	& \Psi_{f_{ij}}^{ij} \\
$Boson$		&\Phi_{IJ}^{\bf{6}} &	& \Phi_{f_{ij}}^{ij} 
		{\rm~~~~~~~where~} I,J=1,...,n; f_{ij}=1,...,a_{ij}^{{\cal R}={\bf 4},{\bf 6}}\\
			&&&~~~~~~~~~~~~~~~~~~~~~~~~~~~~~~~~~~~~~~{\cal R}\otimes 
			{\bf r}_{i}=\bigoplus\limits_{j}a_{ij}^{{\cal R}}\\
\hline
\end{array}
\]
Let us briefly explain what the above table summarises.
In the parent theory,
there are, as mentioned above, gauge bosons $A_{IJ=1,...,n}$ as singlets of $Spin(6)$,
adjoint Weyl fermions $\Psi_{IJ}^{\bf{4}}$
in the fundamental $\bf{4}$ of $SU(4)$ and adjoint scalars 
$\Phi_{IJ}^{\bf{6}}$ in the antisymmetric $\bf{6}$ of $SU(4)$.
The projection is the condition that
\[
A = \gamma(\Gamma) \cdot A \cdot \gamma(\Gamma)^{-1}
\]
for the
gauge bosons and 
\[
\Psi({\rm~or~}\Phi) = R(\Gamma) \cdot \gamma(\Gamma) \cdot 
\Psi({\rm~or~}\Phi) \cdot \gamma(\Gamma)^{-1}
\]
for the fermions and bosons respectively
($\gamma$ and $R$ are appropriate representations of $\Gamma$).

Solving these relations by using Schur's Lemma gives the information on
the orbifold theory.
The equation for $A$ tell us that the original $U(n)$ 
gauge group is broken to
$\prod\limits_{i} SU(N_i)$ where $N_i$ are positive integers such that
$\sum\limits_{i} N_i \dim{\bf r}_i = n$. We point out here that henceforth
we shall use the {\it regular representation} where $n = N|\Gamma|$ for some
integer $N$ and $n_i = N \dim{\bf r}_i$. Indeed other choices are possible
and they give rise to {\it Fractional Branes}, which not only provide interesting
dynamics but are also crucial in showing the equivalence between brane setups
and geometrical engineering \cite{Dog1,Karch1}.
The equations for $\Psi$ and $\Phi$ dictate that they become bi-fundamentals
which transform
under various pairs $(N_i,\bar{N_j})$ within the product gauge group. We have
a total of
$a_{ij}^{\bf{4}}$ Weyl fermions $\Psi _{f_{ij}=1,...,a_{ij}^{\bf{4}}}^{ij}$ 
and $a_{ij}^{\bf 6}$ scalars $\Phi _{f_{ij}}^{ij}$
where $a_{ij}^{\cal R}$ is defined by
\begin{equation}
{\cal R}\otimes {\bf r}_i=\bigoplus\limits_{j}a_{ij}^{\cal R} {\bf r}_j
\label{aij}
\end{equation}
respectively for ${\cal R} = 4,6$.

The supersymmetry of the orbifold theory is determined by analysing the 
commutant of $\Gamma$ as it embeds into the parent $SU(4)$ R-symmetry.
For $\Gamma$ belonging to $SU(2)$, $SU(3)$ or the full $SU(4)$,
we respectively obtain ${\cal N}=2,1,0$. The corresponding geometric
singularities are as presented in the table.
Furthermore, the action of $\Gamma$ clearly differs for $\Gamma \subset
SU(2,3,$~or~$4)$ and the {\bf 4} and {\bf 6} that give rise to
the bi-fundamentals must be decomposed appropriately.
Generically, the number of trivial (principal) 1-dimensional irreducible representations 
corresponds to the co-dimension of the singularity. 
For the matter matrices $a_{ij}$, these irreducible representations give
a contribution of $\delta_{ij}$ and therefore to guaranteed adjoints.
For example, in the case of ${\cal N}=2$, there are
2 trivial {\bf 1}'s in the {\bf 4} and for ${\cal N}=1$, 
${\bf 4} = {\bf 1}_{\rm trivial} \oplus {\bf 3}$.
In our paper, we focus on the latter case since \G
is in $SU(3)$ and gives rise to ${\cal N}=1$. Furthermore we acknowledge the
inherent existence of the trivial 1-dimensional irrep and focus on the decomposition
of the {\bf 3}.

The matrices $a_{ij}^{{\cal R}={\bf 4,6}}$ in (\ref{aij})
and the numbers $\dim{\bf r}_i$ contain
all the information about the matter fields and gauge groups of the orbifold theory.
They can be conveniently encoded into so-called {\bf quiver diagrams}.
Each node of such a diagram treated as a finite graph represents 
a factor in the product gauge group
and is labeled by $\dim{\bf r}_i$. The (possibly oriented) 
adjacency matrix for the graph is prescribed precisely by $a_{ij}$. 
The cases of ${\cal N} = 2,3$ are done \cite{Quiver,Han-He,Muto1,Greene}
and works toward the (non-supersymmetric) ${\cal N} =0$ case are underway \cite{Su4}.
In the ${\cal N} = 2$ case, the quivers must 
coincide with $ADE$ Dynkin diagrams treated
as unoriented graphs in order that the orbifold theory be finite \cite{Mirror}.
The quiver diagrams in general are suggested to be related to WZW modular 
invariants \cite{Han-He,He-Song}.

This is a brief review of the construction via geometric methods and it is our
intent now to see how brane configurations reproduce examples thereof.

\subsection{$D_k$ Quivers from Branes}\label{subsec:Kapustin}
Let us first digress briefly to $A_k$ quivers from branes.
In the case of $SU(2) \supset \Gamma = \widehat{A_k} \simeq Z_{k+1}$, the quiver
theory should be represented by an affine $A_k$ Dynkin diagram, i.e., a regular
polygon with $k+1$ vertices. The gauge group is 
$\prod\limits_{i} SU(N_i) \times U(1)$ 
with $N_i$ being a $k+1$-partition of $n$ since ${\bf r}_i$ are all 
one-dimensional\footnote{The $U(1)$ corresponds to the
centre-of-mass motion and decouples from other parts of the theory 
so that when we discuss
the dynamical properties, it does not contribute.}.
However, we point out that on a classical level we expect
$U(N_i)$'s from the brane perspective rather than $SU(N_i)$. It is only after 
considering the one-loop quantum corrections in the field theory
(or bending in the brane picture) that we realise that
the $U(1)$ factors are frozen. This is explained in \cite{Mlift}.
On the other hand, from the point of view of D-branes
as probes on the orbifold singularity, 
associated to the anomalous $U(1)$'s are field-dependent
Fayet-Illiopoulos terms generating which freezes the $U(1)$ factors.
Thess two prespectives are T-dual to each other.
Further details can be found in \cite{LRA}.

\begin{figure}
\centerline{\psfig{figure=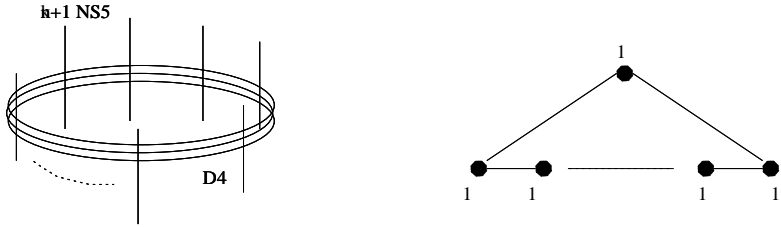,width=4.0in}}
\caption{The ${\cal N}=2$ elliptic model of D4 branes stretched between NS5 branes
	to give quiver theories of the $\widehat{A_k}$ type.}
\label{fig:A}
\end{figure}

Now, placing $k+1$ NS5 branes on a circle with $N_i$ stacked D4 branes 
stretched between the
$i$th and $i+1$st NS5 reproduces precisely this gauge group
with the correct bifundamentals provided by open strings ending on the adjacent 
D4 branes (in the compact direction). This circular model thus furnishes the brane
configuration of an $A_n$-type orbifold theory and is summarised in \fref{fig:A}.
Indeed T-duality in the compact direction transforms the $k+1$ NS5 branes into
a nontrivial metric, viz., the $k+1$-centered Taub-NUT, precisely that expected
from the orbifold picture.
Since both the NS5 and the D4 are offsprings of the M5 brane, in the M-Theory context,
the circular configuration becomes $\R^4 \times \bar\Sigma$ in $\R^{10,1}$, where
$\bar\Sigma$ is a $k+1$-point compactification of a the Riemann surface $\Sigma$
swept out by the worldvolume of the fivebrane \cite{Mlift}. The duality group, which is the
group of automorphisms among the marginal couplings that arise in the resulting field theory,
whence becomes the fundamental group of ${\cal M}_{k+1}$, 
the moduli space of an elliptic curve with $k+1$ marked points.

The introduction of ON$^0$ planes facilitates the next type of ${\cal N}=2,d=4$
quiver theories, namely those encoded by affine $\widehat{D_k}$ Dynkin 
diagrams \cite{Kapustin}. 
The gauge group is now
$SU(2N)^{k-3} \times SU(N)^4 \times U(1)$ (here $U(1)$ decouples also, as explained before) 
dictated by the Dynkin indices of
the $\widehat{D_k}$ diagrams.

\begin{figure}
\centerline{\psfig{figure=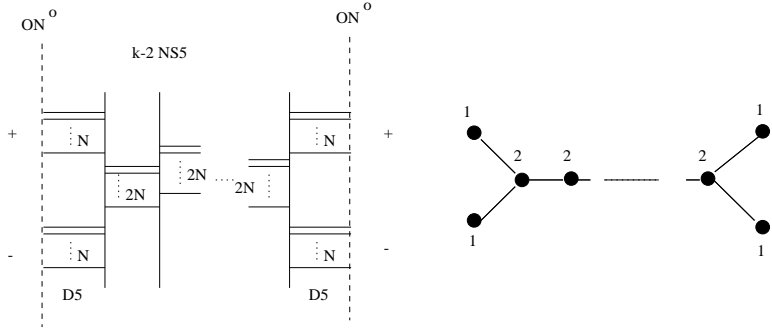,width=4.0in}}
\caption{D5 branes stretched between ON$^0$ branes, interrupted by NS5 branes
	to give quiver theories of the $\widehat{D_k}$ type.}
\label{fig:D}
\end{figure}

There are two ways to see the $\widehat{D_k}$ quiver in the brane picture: one in 
Type IIA theory and the other, in Type IIB. 
Because later on in the construction of the Brane Box 
Model we will use D5 branes which are in Type IIB, we will focus on Type IIB only (for 
a complete description and how the two descriptions are related by T-duality, 
see \cite{Kapustin}).
In this case, what we need is the ON$^0$-plane 
which is the S-dual of a peculiar pair: a D5 brane on top of an O5$^-$-plane.
The one
important property of the ON$^0$-plane is that it has an orbifold description
$\R^6 \times \R^4/{\cal I}$ where
${\cal I}$ is a product of world sheet fermion operator $(-1)^{F_L}$ with the
parity inversion of the $\R^4$ \cite{Sen}.
Let us place 2 parallel vertical ON$^0$ planes and $k-2$ NS5 branes in between and 
parallel to both as in \fref{fig:D}. Between the ON$^0$ and its immediately
adjacent NS5, we stretch $2N$ D5 branes; $N$ of positive charge
on the top and $N$ of negative charge below. 
Now due to the projection of the ON$^0$ plane, $N$ D5 branes of positive charge give
one $SU(N)$ gauge group and $N$ D5 branes of negative charge give
another. Furthermore, these D5 branes end on NS5 branes and the
boundary condition on the NS5 projects out the bi-fundamental hypermultiplets
of these two $SU(N)$ gauge groups 
(for the rules of such projections see \cite{Kapustin}). 
Moreover, between the two adjacent interior
NS5's we stretch $2N$ D5 branes, giving $SU(2N)$'s for the gauge group.
From this brane setup we immediately see 
that the gauge theory is encoded in the affine Quiver diagram of
$\widehat{D_k}$.

\subsection{Brane Boxes}\label{subsec:BBZZ}
We have seen in the last section, that positioning appropriate branes 
according to Dynkin diagrams - which for $\Gamma \subset SU(2)$ have
their adjacency matrices determined by the representation of $\Gamma$,
due to the McKay Correspondence \cite{Han-He} - branengineers some
orbifold theories that can be geometrically engineered. The exceptional
groups however, have so far been elusive
\cite{Kapustin}. For $\Gamma \subset SU(3)$, perhaps related to the fact
that there is not yet a general McKay Correspondence\footnote{For
Gorenstein singularities of dimension 3, only those of the Abelian
type such that 1 is not an eigenvalue of $g$ $\forall g\in \Gamma$ are isolated. 
This restriction perhaps limits na\"{\i}ve brane box
constructions to Abelian orbifold groups \cite{Han-Ura}. For a discussion on
the McKay Correspondence as a ubiquitous thread, see \cite{He-Song}.}
above dimension 2, the problem becomes more subtle; brane setups have 
been achieved for orbifolds of the Abelian type, a restriction that has been
argued to be necessary for consistency \cite{Han-Zaf,Han-Ura}. It is thus
the purpose of this writing to show how a group-theoretic ``twisting''
can relax this condition and move beyond Abelian theories; to this we shall
turn later.

We here briefly review the so-called $Z_k \times Z_{k'}$ elliptic brane
box model. The orbifold theory corresponds to 
$\C^3 / \{ \Gamma = Z_k \times Z_{k'} \subset SU(3) \}$
and hence by arguments before we are
in the realm of ${\cal N} = 1$ super-Yang-Mills. The generators for $\Gamma$
are given, in its fundamental 3-dimensional representation\footnote{We
have chosen the directions in the transverse spacetime upon which
each cyclic factor acts; the choice is arbitrary. In the language of
finite groups, we have chosen the transitivity of the collineation sets.
The group at hand, $Z_k \times Z_{k'}$, is in fact the first example of
an intransitive subgroup of $SU(3)$.
For a discussion of finite subgroups of unitary groups, see \cite{Su4} and
references therein.}, by diagonal
matrices $diag(e^{\frac{2\pi i}{k}},e^{\frac{-2\pi i}{k}},1)$ corresponding
to the $Z_k$ which act non-trivially on the first two coordinates of $\C^3$
and $diag(1,e^{\frac{2\pi i}{k'}},e^{\frac{-2\pi i}{k'}})$
corresponding to the $Z_{k'}$ which act non-trivially on the 
last two coordinates of $\C^3$.

\begin{figure}
\centerline{\psfig{figure=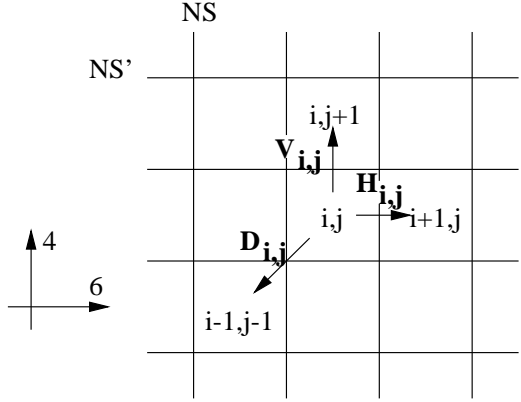,width=2.7in}}
\caption{Bi-fundamentals arising from D5 branes stretched between grids of
	NS5 and NS5$'$ branes in the elliptic brane box model.}
\label{fig:BBZZ}
\end{figure}

Since $\Gamma$ is a direct product of Abelian groups, the representation
thereof is simply a Kronecker tensor product of the two cyclic groups.
Or, from the branes perspective, we should in a sense take a Cartesian 
product or sewing between two ${\cal N}=2$ elliptic $A_{k-1}$ and $A_{k'-1}$ 
models discussed above,
resulting in a brane configuration on $S^1 \times S^1 = T^2$. This is
the essence of the (${\cal N}=1$ elliptic) Brane Box Model \cite{Han-Zaf,Han-Ura}.
Indeed the placement of a perpendicular set of branes breaks the supersymmetry
of the ${\cal N} = 2$ model by one more half, thereby giving the desired
${\cal N}=1$. More specifically, we place $k$ NS5 branes in the $012345$
and $k'$ NS5$'$ branes in the $012367$ directions, whereby forming
a grid of $kk'$ boxes as in \fref{fig:BBZZ}.
We then stretch $n_{ij}$ D5 branes in the $012346$ directions
within the $i,j$-th box and compactify the $46$ directions (thus making
the low-energy theory on the D5 brane to be 4 dimensional).
The bi-fundamental fields are then given according to adjacent boxes
horizontally, vertically and diagonally and the gauge groups is
$(\bigotimes\limits_{i,j}SU(N))\times U(1)  = SU(N)^{kk'}\times U(1)$ (here
again the $U(1)$ decouples) as expected from geometric
methods. Essentially we construct one box for each irreducible representation of 
$\Gamma=Z_k \times Z_{k'}$ such that going in the 3 directions as shown
in \fref{fig:BBZZ} corresponds to tensor decomposition of the irreducible
representation in that grid and a special ${\bf 3}$-dimension representation
which we choose when we construct the Brane Box Model.

We therefore see the realisation of Abelian orbifold theories in dimension
3 as brane box configurations; twisted identifications of the grid can
in fact lead to more exotic groups such as $Z_k \times Z_{kk'/l}$.
More details can be found in \cite{Han-Ura}.

\section{The Group $G=Z_k \times D_{k'}$} \label{sec:group}
It is our intent now to investigate the next simplest example of intransitive
subgroups of $SU(3)$, i.e., the next infinite series of orbifold theories
in dimension 3 (For definitions on the classification of collineation groups,
see for example \cite{Su4}). This will give us a first example of a Brane Box Model
that corresponds to non-Abelian singularities.

Motivated by the $Z_k \times Z_{k'}$ treated in section
\sref{sec:review}, we let the second factor be the binary dihedral group
of $SU(2)$, or the $D_{k'}$ series (we must point out that in our notation, 
the $D_{k'}$ group gives the $\widehat{D}_{k'+2}$ Dynkin diagram). 
Therefore $\Gamma$ is the group
$G = Z_k \times D_{k'}$, generated by
\[
\alpha = \left(  \begin{array}{ccc}
                        \omega_{k} & 0 & 0  \\
                         0 & \omega_{k}^{-1} & 0\\
                         0  &  0 & 1
                \end{array}
        \right)
~~~~~~~~
\beta = \left(  \begin{array}{ccc}
                        1  & 0  & 0  \\
                        0 & \omega_{2k'} & 0  \\
                        0 &  0 & \omega_{2k'}^{-1}  
                \end{array}
        \right)
~~~~~~~~
\gamma =\left(  \begin{array}{ccc} 
                1  &  0  &  0  \\
                0  &  0  &  i \\
                0  &  i &   0 
                \end{array}
        \right)
\]
where $w_x := e^{\frac{2 \pi i}{x}}$. We observe that indeed $\alpha$ generates
the $Z_k$ acting on the first two directions in $\C^3$ while $\beta$ and $\gamma$
generate the $D_{k'}$ acting on the second two.

We now present some crucial properties of this group $G$ which shall be used in
the next section. First we remark that the $\times$ in $G$ is really an abuse of
notation, since $G$ is certainly not a direct product of these two groups. This
is the cause why na\"{\i}ve constructions of the Brane Box Model fail and to this
point we shall turn later.
What we really mean is that the actions on the first two and last two coordinates
in the transverse directions by these subgroups are to be construed as separate.
Abstractly, we can write the presentation of $G$ as
\begin{equation}
\alpha \beta = \beta \alpha,~~~~\beta \gamma =\gamma \beta^{-1},~~~~
\alpha^{m} \gamma \alpha^{n} \gamma =\gamma \alpha^{n} \gamma \alpha^{m}
~~~~\forall m,n \in \Z
\label{relations}
\end{equation}
These relations compel all elements in $G$ to be writable in the form
$\alpha^{m} \gamma \alpha^{\tilde{m}} \gamma^{n} \beta^{p}$. However, before discussing
the whole group, we find it very useful to discuss the subgroup generated by $\beta$ and
$\gamma$, i.e the binary dihedral group $D_{k'}$ as a degenerate ($k=1$) case of $G$,
because the properties of the binary dihedral group turn out to be crucial for the structure
of the Brane Box Model and the meaning of ``twisting'' which we shall clarify later.

\subsection{The Binary Dihedral $D_{k'} \subset G$}\label{subsec:D}
All the elements of $D_{k'}$ can be written as
$ \beta^{p}\gamma^{n}$ with $n=0,1$ and $p=0,1,...,2k'-1$, giving the
order of the group as $4k'$. We now move onto Frobenius characters.
It is easy to work out the structure of conjugate classes. We have two conjugate
classes $(1), (\beta^{k'})$ which have only one element, $(k'-1)$ conjugate classes
$(\beta^p,\beta^{-p}),p=1,..,k'-1$ which have two elements and two conjugate classes
$( \beta^{p~\even}\gamma),( \beta^{p~\odd}\gamma)$ which have $k'$ elements.
The class equation is thus as follows:
\[
4k' = 1 + 1 + (k' - 1)\cdot 2 + 2 \cdot k'.
\]
Moreover there are 4 1-dimensional and $k'-1$ 2-dimensional irreducible
representations
such that the characters for the 1-dimensionals depend on the parity of $k'$.
Now we have enough facts to clarify our notation: the group $D_{k'}$ gives
$k'+3$ nodes (irreducible representations) which corresponds to the Dynkin diagram of 
$\widehat{D_{k'+2}}$.

We summarise the character table as follows:\\
\[
\doublerulesep 0.7pt
\begin{array}{cc}
k' \even &
\begin{array}{|c|c|c|c|c|c|c|}
\hline \hline
        & C_{n=0}^{p=0} &  C_{n=0}^{p=k'} &  C_{n=0}^{\pm \even~p} &
	  C_{n=0}^{\pm \odd~p} & C_{n=1}^{\even~p} & C_{n=1}^{\odd~p} \\ \hline
|C|  	& 1 & 1 & 2 & 2  & k' & k' \\ \hline
\#C   	& 1 & 1 & \frac{k'-1}{2} & \frac{k'-1}{2} 
	& 1 & 1 \\ \hline \hline
\Gamma_1	& 1 &  1 &  1 &  1  & 1 & 1 \\ \hline
\Gamma_2	& 1 &  -1 &  1 &  -1 & 1 & -1 \\ \hline
\Gamma_3	& 1 & 1 & 1 & 1 & -1 & -1 \\ \hline
\Gamma_4	& 1 & -1 & 1 & -1 & -1 & 1 \\ \hline
\Gamma_l	& \multicolumn{4}{|c|}{(\omega_{2k'}^{lp}+
			\omega_{2k'}^{-lp})~~~~l=1,..,k'-1} & 0 & 0 \\ \hline
\end{array}
\\ \\
k' \odd &
\doublerulesep 0.7pt
\begin{array}{|c|c|c|c|c|c|c|}
\hline \hline
        & C_{n=0}^{p=0} &  C_{n=0}^{p=k'} &  C_{n=0}^{\pm \even~p} &
	  C_{n=0}^{\pm \odd~p} & C_{n=1}^{\even~p} & C_{n=1}^{\odd~p} \\ \hline
|C|  	& 1 & 1 & 2 & 2  & k' & k' \\ \hline
\#C   	& 1 & 1 & \frac{k'-2}{2} & \frac{k'}{2} 
	& 1 & 1 \\ \hline \hline
\Gamma_1	& 1 & 1 &  1 &  1  & 1 & 1 \\ \hline
\Gamma_2	& 1 & 1 &  1 &  -1 & \omega_4 & -\omega_4 \\ \hline
\Gamma_3	& 1 & 1 & 1 & 1 & -1 & -1 \\ \hline
\Gamma_4	& 1 & 1 & 1 & -1 & -\omega_4 & \omega_4 \\ \hline
\Gamma_l	& \multicolumn{4}{|c|}{(\omega_{2k'}^{lp}+
			\omega_{2k'}^{-lp})~~~~l=1,..,k'-1} & 0 & 0 \\ \hline
\end{array}
\\
\end{array}
\]

In the above tables, $|C|$ denotes the number of group elements in conjugate class
$C$ and $\#C$, the number of conjugate classes belonging to this type. Therefore
$\sum\limits_C \#C\cdot|C|$ should equal to order of the group.
When we try to
look for the character of the 1-dimensional irreps, we find it to be the same as the
character of the factor group $D_{k'}/N$ where $N$ is the normal subgroup generated by 
$\beta$. This factor group is Abelian of order 4 and is different depending
on the parity of $k'$. When $k'=\even$, it
is $Z_2\times Z_2$ and when $k'=\odd$ it is $Z_4$. Furthermore, the conjugate class
$(\beta^p,\beta^{-p})$ corresponds to different elements in this factor group 
depending on the parity of $p$, and we distinguish the two different cases in the 
table as $C_{n=0}^{\pm \odd~p}$ and $C_{n=0}^{\pm \even~p}$.

\subsection{The whole group $G = Z_k \times D_{k'}$}
Now from (\ref{relations}) we see that all elements of G can be written in the form
$\alpha^{m} \gamma \alpha^{\tilde{m}} \gamma^{n} \beta^{p}$ with 
$m,\tilde{m}=0,..,k-1$, $n=0,1$ and $p=0,..2k'-1$, which we abbreviate as
$(m,\tilde{m},n,p)$. In the matrix form of our fundamental representation, they become
\[
\begin{array}{ll}
(m,\tilde{m},n=0,p)= & (m,\tilde{m},n=1,p)= \\

\left(  \begin{array}{ccc}
                        \omega_{k}^{m+\tilde{m}} & 0 & 0\\
                        0 & 0 & i\omega_{k}^{-m} \omega_{2k'}^{-p}  \\
                        0 & i\omega_{k}^{-\tilde{m}}\omega_{2k'}^{p} & 0
                        \end{array}
                \right),
&
\left(  \begin{array}{ccc}
                        \omega_{k}^{m+\tilde{m}} & 0 & 0 \\
                         0 & -\omega_{k}^{-m} \omega_{2k'}^{p} & 0 \\
                        0 & 0 & -\omega_{k}^{-\tilde{m}}\omega_{2k'}^{-p} 
                        \end{array}
                \right). \\
\end{array}
\]
Of course this representation is not faithful and there is a non-trivial orbit; we can
easily check the repeats:
\begin{equation}
\begin{array}{l}
(m,\tilde{m},n=0,p)=(m+\frac{k}{(k,2k')},\tilde{m}-\frac{k}{(k,2k')},n=0,p-\frac{2k'}
{(k,2k')}), \\
(m,\tilde{m},n=1,p)=(m+\frac{k}{(k,2k')},\tilde{m}-\frac{k}{(k,2k')},n=1,p+\frac{2k'}
{(k,2k')}) 
\end{array}
\label{orbit}
\end{equation}
where $(k,2k')$ denotes the largest common divisor between them. Dividing by the
factor of this repeat immediately gives the 
order of $G$ to be $\frac{4k'k^2}{(k,2k')}$.

We now move on to the study of the characters of the group. The details of the
conjugation automorphism, class equation and irreducible representations we shall leave to the Appendix
and the character tables we shall present below; again we have two cases, 
depending on the parity of $\frac{2k'}{(k,2k')}$. First however we start with
some preliminary definitions. We define $\eta$ as a function of $n$, $p$ and
$h = 1,2,3,4$.
\begin{equation}
\begin{array}{cc}
k' = \even &
\begin{array}{ccccc}
     &  (n=1,p=\even) & (n=1,p=\odd) & (n=0,p=\even) & (n=0,p=\odd) \\
\eta^{1} & 1 & 1 & 1 & 1 \\
\eta^{2} & 1 & -1 & 1 & -1 \\
\eta^{3} & 1 & 1 & -1 & -1 \\
\eta^{4} & 1 & -1 & -1 & 1 
\end{array}
\\
k' = \odd &
\begin{array}{ccccc}
        & (n=1,p=\odd) & (n=1,p=\even) & (n=0,p=\even) & (n=0,p=\odd)\\
\eta^{1} & 1 & 1 & 1 & 1 \\
\eta^{2} & 1 & -1 & \omega_4 & -\omega_4 \\
\eta^{3} & 1 & 1 & -1 & -1 \\
\eta^{4} & 1 & -1 & - \omega_4 &  \omega_4
\end{array}
\\
\end{array}
\label{eta}
\end{equation}
Those two tables simply give the character tables of $Z_2\times Z_2$ and $Z_4$
which we saw in the last section.

Henceforth we define $\delta := (k,2k')$. 
Furthermore, we shall let 
$\Gamma^n_x$ denote an $n$-dimensional irreducible representation indexed by some (multi-index) $x$.
For $\frac{2k'}{\delta} = \even$,
there are $4k$ 1-dimensional irreducible representations indexed by $(l,h)$ with $l=0,1,..,k-1$ and 
$h=1,2,3,4$ and
$k(\frac{k'k}{(k,2k')}-1)$ 2-dimensionals indexed by $(d,l)$ with
$d=1,..,\frac{k'k}{(k,2k')}-1;l=0,..,k-1$.
For $\frac{2k'}{\delta} = \odd$,
there are $2k$ 1-dimensional irreducible representations indexed by $(l,h)$ with $l=0,..,k-1;h=1,3$ and
$k(\frac{k'k}{(k,2k')}-\frac{1}{2})$ 2-dimensionals indexed by $(d,l)$
$d=1,..,\frac{k'k}{(k,2k')}-1;l=0,..,k-1$ and
$d=\frac{k'k}{(k,2k')};l=0,..,\frac{k}{2}-1$.
Now we present the character tables.\\

{\large $\frac{2k'}{\delta} = \even$}
\[
\doublerulesep 0.7pt
\begin{array}{|c|c|c|c|}
\hline \hline
|C| & 1 & 2 & \frac{k'k}{(k,2k')} \\ \hline
\#C & 2k & k(\frac{k'k}{(k,2k')}-1) & 2k \\
\hline \hline
	&
	\begin{array}{c}
		m = 0,..,\frac{k}{\delta}-1; ~i = 0,..,\delta-1; \\
			~\tilde{m} = m + \frac{i k}{\delta}; ~n=1; \\
			~p=k'-\frac{ik'}{(k,2k')},2k'-\frac{ik'}{(k,2k')}
	\end{array}
	&
	\begin{array}{c}
		m = 0,..,\frac{k}{\delta}-1; ~i = 0,..,\delta-1; ~n=1 \\
		\left\{
		\begin{array}{l}
	  	s = 0,..,m-1; ~p = 0,..2k'-1;\\
			~~~~~~\tilde{m} = s + \frac{i k}{\delta};\\
	  	s = m; \mbox{and require further that} \\
			~~~~~~p < (-p - \frac{2 i k'}{\delta}) \bmod (2k')\\
		\end{array}
		\right.
	\end{array}
	&
	\begin{array}{c}
		m = 0;\\
		\tilde{m} = 0,..,k-1;\\
		p = 0,1;\\
		n = 0\\
	\end{array} \\ \hline
\Gamma^1_{(l,h)} & \multicolumn{3}{|c|}{
	\omega_{k}^{(m+\tilde{m})l} \eta^{h},~~~~~~l=0,1,..,k-1;~~h=1,..,4}
	\\ \hline
\Gamma^2_{(d,l)} & \multicolumn{2}{|c|}{
	\begin{array}{c}
	(-1)^{d}(\omega_{k}^{-dm}\omega_{2k'}^{dp}+\omega_{k}^{-d\tilde{m}}
	\omega_{2k'}^{-dp}) \omega_{k}^{(m+\tilde{m})l}\\
	~~~~~~~~~~~~~d\in[1,\frac{k'k}{(k,2k')}-1];~~l\in [0,k)\\
	\end{array}
	}
	& 0 \\ \hline
\end{array}
\]

{\large $\frac{2k'}{\delta} = \odd$}
\[
\doublerulesep 0.7pt
\begin{array}{|c|c|c|c|}
\hline \hline
|C| & 1 & 2 & \frac{k'k}{(k,2k')} \\ \hline
\#C & k & k(\frac{k'k}{(k,2k')}-\frac12) & k \\
\hline \hline
	&
	\begin{array}{c}
		m = 0,..,\frac{k}{\delta}-1;\\
		i = 0,..,\delta-1 \mbox{ and even}; \\
			~\tilde{m} = m + \frac{i k}{\delta}; ~n=1; \\
			~p=k'-\frac{ik'}{(k,2k')},\\~~~~~~2k'-\frac{ik'}{(k,2k')}
	\end{array}
	&
	\begin{array}{c}
		m = 0,..,\frac{k}{\delta}-1; ~i = 0,..,\delta-1; ~n=1 \\
		\left\{
		\begin{array}{l}
	  	s = 0,..,m-1; ~p = 0,..2k'-1;\\
			~~~~~~\tilde{m} = s + \frac{i k}{\delta};\\
	  	s = m; \mbox{and require further that} \\
			~~~~~~p < (-p - \frac{2 i k'}{\delta}) \bmod (2k') 
				\mbox{ for even } i\\
			~~~~~~p \le (-p - \frac{2 i k'}{\delta}) \bmod (2k') 
				\mbox{ for odd } i\\	
		\end{array}
		\right.
	\end{array}
	&
	\begin{array}{c}
		m = 0;\\
		\tilde{m} = 0,..,k-1;\\
		p = 0;\\
		n = 0\\
	\end{array} \\ \hline
\Gamma^1_{(l,h)} & \multicolumn{3}{|c|}{
	\omega_{k}^{(m+\tilde{m})l} \eta^{h},~~~~~~l=0,1,..,k-1;~~h=1,3}
	\\ \hline
\Gamma^2_{(d,l)} & \multicolumn{2}{|c|}{
	\begin{array}{c}
	(-1)^{d}(\omega_{k}^{-dm}\omega_{2k'}^{dp}+\omega_{k}^{-d\tilde{m}}
	\omega_{2k'}^{-dp}) \omega_{k}^{(m+\tilde{m})l}\\
	~~~~~~~~~~~~d\in[1,\frac{k'k}{(k,2k')}-1];~~l\in [0,k)\\
	\end{array}
	}
	& 0 \\ \hline
\Gamma^2_{(d,l)} & \multicolumn{2}{|c|}{
	\begin{array}{c}
	(-1)^{d}(\omega_{k}^{-dm}\omega_{2k'}^{dp}+\omega_{k}^{-d\tilde{m}}
	\omega_{2k'}^{-dp}) \omega_{k}^{(m+\tilde{m})l}\\
	~~~~~~~~~~~~d=\frac{k'k}{(k,2k')};~~l\in [0,\frac{k}{2})\\
	\end{array}
	}
	& 0 \\ \hline
\end{array}
\]

Let us explain the above tables in more detail. The third row of each table
give the representative elements of the various conjugate classes. 
The detailed description of the group elements in 
each conjugacy class is given in the Appendix.
It is easy to see, by using the above character tables, that given two
elements $(m_i,\tilde{m}_i,n_i,p_i)~~i=1,2$, if they share the same 
characters (as given in the last two rows), they belong to 
same conjugate class as to be expected since the character is a class
function.

We can be more precise and actually write down the 2 dimensional irreducible representation
indexed by $(d,l)$ as
\begin{equation}
\begin{array}{l}
(m,\tilde{m},n=0,p)= \omega_{k}^{(m+\tilde{m})l} 
\left(   \begin{array}{cc}
0 & i^d \omega_{k}^{-dm} \omega_{2k'}^{-dp} \\
i^d \omega_{k}^{-d\tilde{m}} \omega_{2k'}^{dp} & 0
\end{array} \right)
\\
(m,\tilde{m},n=1,p)=\omega_{k}^{(m+\tilde{m})l} 
\left(   \begin{array}{cc}
 (-1)^d \omega_{k}^{-dm} \omega_{2k'}^{dp} & 0 \\
0 & (-1)^d \omega_{k}^{-d\tilde{m}} \omega_{2k'}^{-dp}
\end{array} \right)
\end{array}
\label{2d}
\end{equation}

\subsection{The Tensor Product Decomposition in $G$}\label{subsec:decomp}
A concept crucial to character theory and representations 
is the decomposition of tensor products into tensor sums among the
various irreducible representations, namely the equation
\[
{\bf r}_k \otimes {\bf r}_i = \bigoplus\limits_j a_{ij}^k {\bf r}_j.
\]
Not only will such an equation enlighten us as to the structure of the
group, it will also provide quintessential information to the brane box
construction to which we shall turn later. Indeed the ${\cal R}$ in
(\ref{aij}) is decomposed into direct sums of irreducible representations ${\bf r}_k$, which
by the additive property of the characters, makes the fermionic and bosonic
matter matrices $a_{ij}^{\cal R}$ ordinary sums of matrices $a_{ij}^k$.
In particular, knowing the specific decomposition of the {\bf 3},
we can immediately construct the quiver diagram prescribed by 
$a_{ij}^{\bf 3}$ as discussed in \sref{subsec:Quiver}.

We summarise the decomposition laws as follows (using the multi-index notation
for the irreducible representations introduced in the previous section).

{\large $\frac{2k'}{\delta} = \even$}
\[
\begin{array}{|c|c|}
\hline
{\bf 1} \otimes {\bf 1}' & (l_1,h_1)_1 \otimes (l_2,h_2)_1 = (l_1+l_2,h_3)_1\\
		& \mbox{where $h_3$ is such that $\eta^{h_1}\eta^{h_2} = \eta^{h_3}$
		according to (\ref{eta}).} \\ \hline
{\bf 2} \otimes {\bf 1}  & (d,l_1)_2 \otimes (l_2,h_2)_1=\left\{
	\begin{array}{l}
        	(d,l_1+l_2)_2~~{\rm when}~~h_2=1,3.  \\
        	(\frac{k'k}{(k,2k')}-d,l_1+l_2-d)_2~~{\rm when}~~h_2=2,4
        \end{array}
        \right.
	\\ \hline
{\bf 2} \otimes {\bf 2}' &
	\begin{array}{l}
	(d_1,l_1)_2 \otimes (d_2 \le d_1,l_2)_2 = \\
	~~~~~~~~~~(d_1+d_2,l_1+l_2)_2 \oplus (d_1-d_2,l_1+l_2-d_2)_2 \\
	{\rm where} \\
	(d_1-d_2,l_1+l_2-d_2)_2 := \\
	~~~~~~~~~~(l_1+l_2-d_2,h=1)_1 \oplus (l_1+l_2-d_2,h=3)_1 
	{\rm~~if~~} d_1 = d_2 \\
	(d_1+d_2,l_1+l_2)_2 := \\
	~~~~~~~~~~(l_1+l_2,h=2)_1 \oplus (l_1+l_2,h=4)_1
	{\rm~~if~~} d_1+d_2=\frac{k'k}{\delta} \\
	(d_1+d_2,l_1+l_2)_2 := \\
	~~~~~~~~~~(\frac{2k'k}{(k,2k')}-(d_1+d_2),(l_1+l_2)-(d_1+d_2))_2
	{\rm~~if~~} d_1+d_2>\frac{k'k}{\delta} \\
	\end{array}
	\\ \hline
\end{array}
\]

{\large $\frac{2k'}{\delta} = \odd$}
\[
\begin{array}{|c|c|}
\hline
{\bf 1} \otimes {\bf 1}' & (l_1,h_1)_1 \otimes (l_2,h_2)_1=\left\{
	\begin{array}{l}
	        (l_1+l_2,h=1)_1~~{\rm if}~~h_1=h_2  \\
        	(l_1+l_2,h=3)_1~~{\rm if}~~h_1\neq h_2
	\end{array}   \right.
		\\ \hline
{\bf 2} \otimes {\bf 1}  & (d,l_1)_2 \otimes (l_2,h_2)_1=\left\{
	\begin{array}{l}
		(d,l_1+l_2)_2 \\
		(d,l_1+l_2-\frac{k}{2})_2 {\rm~~if~~} 
		d=\frac{k'k}{(k,2k')} {\rm~and~} l_1+l_2 \ge \frac{k}{2}
	\end{array} \right.
	\\ \hline
{\bf 2} \otimes {\bf 2}' &
	\begin{array}{l}
	(d_1,l_1)_2 \otimes (d_2 \le d_1,l_2)_2 = \\
	~~~~~~~~~~(d_1+d_2,l_1+l_2)_2 \oplus (d_1-d_2,l_1+l_2-d_2)_2 \\
	{\rm where} \\
	(d_1-d_2,l_1+l_2-d_2)_2 := \\
	~~~~~~~~~~(l_1+l_2-d_2,h=1)_1 \oplus (l_1+l_2-d_2,h=3)_1
	{\rm~~if~~} d_1 = d_2 \\
	(d_1+d_2,l_1+l_2)_2 := \\
	~~~~~~~~~~(d_1 + d_2, l_1 + l_2 - \frac{k}{2})_2
	{\rm~~if~~} d_1+d_2=\frac{k'k}{\delta}{\rm~and~} l_1 + l_2 \ge \frac{k}{2}\\
	(d_1+d_2,l_1+l_2)_2 := \\
	~~~~~~~~~~(\frac{2k'k}{(k,2k')}-(d_1+d_2),(l_1+l_2)-(d_1+d_2))_2
	{\rm~~if~~} d_1+d_2>\frac{k'k}{\delta} \\
	\end{array}
	\\ \hline
\end{array}
\]

\subsection{$D_{\frac{kk'}{\delta}}$, an Important Normal Subgroup}\label{subsec:H}
We now investigate a crucial normal subgroup $H \triangleleft G$. The purpose
is to write $G$ as a canonical product of $H$ with the factor group formed by
quotienting $G$ thereby, i.e., as $G \simeq G/H \times H$. 
The need for this rewriting of the group will 
become clear in \sref{sec:BB} on the brane box construction.
The subgroup we desire is the one presented in the following:
\begin{lemma}
The subgroup
\[
H := \{(m,-m,n,p)|m=0,..,k-1;n=0,1;p=0,...,2k'-1 \}
\]
is normal in $G$ and is isomorphic to $D_{\frac{kk'}{\delta}}$.
\end{lemma}
To prove normality we use the multiplication and conjugation rules in $G$ 
given in the Appendix as (\ref{conj}) and (\ref{multi}).
Moreover, let $D_{\frac{kk'}{\delta}}$ be generated by $\tilde{\beta}$ and 
$\tilde{\gamma}$ using the notation of \sref{subsec:D}, then isomorphism
can be shown by the following bijection:
\[
\begin{array}{l}
(m,-m,1,p) \longleftrightarrow \tilde{\beta}^{\frac{2k'}{\delta}m-
\frac{k}{\delta}(p-k')},  \\
(m,-m,0,p) \longleftrightarrow \tilde{\beta}^{\frac{2k'}{\delta}m+
\frac{k}{\delta}p} \tilde{\gamma}.
\end{array}
\]
Another useful fact is the following:
\begin{lemma}
The factor group $G/H$ is isomorphic to $Z_k$.
\end{lemma}
This is seen by noting that $\alpha^l,l=0,1,...k-1$ can be used as representatives
of the cosets. We summarise these results into the following

\begin{proposition}
There exists another representation of $G$, namely
$Z_k \times D_{k'} \simeq Z_k \rtimes D_{\frac{kk'}{\delta}}$, generated by the same
$\alpha$ together with
\[
\begin{array}{cc}
\tilde{\beta}^{\frac{2k'}{\delta}m- \frac{k}{\delta}p} := (m,-m,1,p+k') = ~~~~~~~~~~
 & \tilde{\gamma} := \gamma = (0,0,0,0)= ~~~~~~~~~~\\
\left(  \begin{array}{ccc}
1  &  0  &  0  \\
0 & \omega_{k}^{-m} \omega_{2k'}^{p}  & 0  \\
0 & 0 & \omega_{k}^{m} \omega_{2k'}^{-p}   
\end{array}  \right), &
\left(  \begin{array}{ccc}
1 & 0 & 0 \\
0 & 0 & i \\
0 & i & 0 \\
\end{array}  \right).
\end{array}
\]
The elements of the group can now be written as
$\alpha^a \tilde{\beta}^b \tilde{\gamma}^n$ with $a\in [0,k)$, 
$b\in [0,\frac{2kk'}{\delta})$ and $n=0,1$, constrained by the presentation
\[
\{\alpha^k=\tilde{\beta}^{\frac{2kk'}{\delta}}=1,
\tilde{\beta}^{\frac{kk'}{\delta}}=\tilde{\gamma}^2=-1,
\alpha \tilde{\beta} = \tilde{\beta} \alpha , \tilde{\beta} \tilde{\gamma} =
\tilde{\gamma} \tilde{\beta}^{-1},
\alpha  \tilde{\gamma} =\tilde{\beta}^{\frac{2k'}{\delta}} \tilde{\gamma} \alpha\}
\]
\end{proposition}
In the proposition, by $\rtimes$ we do mean the internal semi-direct product between
$Z_k$ and $H := D_{\tilde{k}} := D_{\frac{kk'}{\delta}}$, in the sense \cite{Alperin} that
(I) $G=HZ_k$ as cosets, (II) $H$ is normal in $G$ and $Z_k$ is another subgroup, 
and (III) $H \cap Z_k = 1$. Now we no longer abuse the symbol $\times$ and
unambiguously use $\rtimes$ to show the true structure of $G$.
We remark that this representation is in some sense more natural (later we shall see
that this naturality is not only mathematical but also physical). The mathematical
natuality is seen by the lift from the normal subgroup
$H$. We will see what is the exact meaning 
of the ``twist'' we have mentioned before. When we include the generator 
$\alpha$ and lift the normal subgroup $D_{\frac{kk'}{\delta}}$ 
to the whole group $G$, the structure of
conjugacy classes will generically change as well. For example, from
\begin{equation}
\alpha (\tilde{\beta}^b \tilde{\gamma}) \alpha^{-1} =
(\tilde{\beta}^{b+\frac{2k'}{\delta}} \tilde{\gamma}),
\label{add1}
\end{equation}
we see that the two different conjugacy classes $(\tilde{\beta}^{\even~b} \tilde{\gamma})$
and $(\tilde{\beta}^{\odd~b} \tilde{\gamma})$ 
will remain distinct if $\frac{2k'}{\delta}=\even$
and collapse into one single conjugacy class 
if $\frac{2k'}{\delta}=\odd$. We formally call the latter case
{\bf twisted}. Further clarifications regarding the 
structure of the conjugacy classes
of $G$ from the new points of view, especially physical, shall be most welcome.

After some algebraic manipulation, we can
write down all the conjugacy classes of $G$ in this new description.
For fixed $a$ and $\frac{2k'}{\delta}=\even$, 
we have the following classes: $(\alpha^a \tilde{\beta}^{-\frac{k'}{\delta}a})$,
$(\alpha^a \tilde{\beta}^{\frac{kk'}{\delta}-\frac{k'}{\delta}a}),$
$(\alpha^a \tilde{\beta}^b, \alpha^a \tilde{\beta}^{-b-\frac{2k'}{\delta}a})$
(with $b \neq -\frac{k'}{\delta}a$ and $\frac{kk'}{\delta}-\frac{k'}{\delta}a$), 
$(\alpha^b \tilde{\beta}^{p~\even} \tilde{\gamma})$ and 
$(\alpha^b \tilde{\beta}^{p~\odd} \tilde{\gamma})$. The crucial point here is that,
for every value of $a$, the structure of conjugacy classes is almost the same as that
of $D_{\frac{kk'}{\delta}}$. There is a 1-1
correspondence (or the lifting without the ``twist'') as we go from the
conjugacy classes of $H$ to $G$, making it possible to use
the idea of \cite{Han-Zaf2} to construct the corresponding
Brane Box Model. We will see this point more clearly later.
On the other hand, when
$\frac{2k'}{\delta}=\odd$, for fixed $a$, the conjugacy classes are 
no longer in 1-1 correspondence
between $H$ and $G$. Firstly, the last two 
classes of $H$ will combine into only one of $G$. Secondly,
the classes which contain only one element (the first two in $H$) will remain
so only for $a=\even$; for $a=\odd$, the they will combine into
one single class of $G$ which has two elements. 

So far the case of $\frac{2k'}{\delta}=\odd$ befuddles us and we do not know 
how the twist obstructs the construction of the Brane Box Model. This
twist seems to suggest quiver theories on non-affine $D_k$ diagrams because
the bifurcation on one side collapses into a single node, a phenomenon
hinted before in \cite{Han-He,Han-Zaf2}.
It is a very interesting problem which we leave to further work.

\section{The Brane Box for \G} \label{sec:BB}
\subsection{The Puzzle}
The astute readers may have by now questioned themselves why such a long digression on
the esoterica of $G$ was done; indeed is it not enough to straightforwardly combine the
$D_{k'}$ quiver technique with the elliptic model and stack $k$ copies of Kapustin's
configuration on a circle to give the \G brane boxes?
Let us investigate where this na\"{\i}vet\'{e} fails. According to the discussions in
\sref{subsec:BBZZ}, one must construct one box for each irreducible representation of $G$. Let us place 2
ON$^0$ planes with $k'$ parallel NS5 branes in between as in \sref{subsec:Kapustin},
and then copy this $k$ times in the direction of the ON$^0$ and compactify that direction.
This would give us $k + k$ boxes each containing 2 1-dimensional irreducible representations corresponding
to the boxes bounded by one ON$^0$ and one NS5 on the two ends. And in the middle
we would have $k(k'-1)$ boxes each containing 1 2-dimensional irreducible representation.

Therefrom arises a paradox already! From the discussion of the group 
$G=Z_k \times D_{k'}$ in \sref{sec:group}, we recall that there are
$4k$ 1-dimensional irreducible representations and $k(\frac{k'k}{(k,2k')}-1)$ 2-dimensionals if
$\frac{2k'}{\delta} = \even$ and for $\frac{2k'}{\delta} = \odd$, $2k$ 
1-dimensionals and $k(\frac{k'k}{(k,2k')}-\frac12)$ 2-dimensionals.
Our attempt above gives a mismatch of the number the 2-dimensionals
by a factor of as large as $k$; there are far too many 2-dimensionals
for $G$ to be placed into the required $kk'$ boxes.
This mismatch tells us that such na\"{\i}ve constructions of the Brane Box Model fails.
The reason is that in this case what we are dealing with is a non-Abelian group
and the noncommutative property thereof twists the na\"{\i}ve structure
of the singularity. To correctly account for the property of the singularity 
after the non-Abelian twisting, we should attack in a new direction.
In fact, the discussion of the normal
subgroup $H$ in \sref{subsec:H} is precisely the way to see the 
structure of singularity more properly.
Indeed we have hinted, at least for $\frac{2k'}{\delta}=\even$,
that the na\"{\i}ve structure of the Brane Box Model can be applied again with a little 
modification, i.e., with the replacement of $D_{k'}$ by $D_{\frac{kk'}{\delta}}$.
Here again we have the generator of $Z_k$ acting on the first two coordinates of
$\C^3$ and the generators of $D_{\frac{kk'}{\delta}}$ acting on the last two.
This is the subject of the next sub section where we will 
give a consistent Brane Box Model for $G = Z_k \times D_{k'}$.

\subsection{The Construction of Brane Box Model}
Let us first discuss the decomposition of the fermionic {\bf 4} for which we
shall construct the brane box (indeed the model will dictate the fermion 
bi-fundamentals, bosonic matter fields will be given therefrom by supersymmetry).
As discussed in \cite{Han-He} and \sref{subsec:Quiver}, 
since we are in an ${\cal N}=1$ (i.e., a co-dimension
one theory in the orbifold picture), the {\bf 4} must decompose into 
${\bf 1} \oplus {\bf 3}$ with the {\bf 1} being trivial. More precisely, 
since $G$ has only 1-dimensional or 2-dimensional irreducible representations, for giving the correct quiver diagram which corresponds to the Brane Box Model
the
{\bf 4} should  go into one trivial 1-dimensional, one non-trivial 1-dimensional
and one 2-dimensional according to
\[
{\bf 4} \longrightarrow (0,1)_1 \oplus (l',h')_1 \oplus (d,l)_2.
\]
Of course we need a constraint so as to ensure that such a decomposition
is consistent with the unity-determinant condition of the matrix representation
of the groups. Since from (\ref{2d}) we can compute the determinant of the $(d,l)_2$ to be
$(-1)^{(n+1)(d+1)}\omega_{k}^{(m+\tilde{m})(2l-d)}$, the constraining condition
is  $l'+2l-d\equiv 0(\bmod k)$.
In particular we choose
\begin{equation}
{\bf 3} \longrightarrow (l'=1,h'=1)_1+(d=1,l=0)_2;
\label{decomp}
\end{equation}
indeed this choice is precisely in accordance with the defining matrices of
$G$ in \sref{sec:group} and we will give the Brane Box Model corresponding to this
decomposition and check consistency.

\begin{figure}
\centerline{\psfig{figure=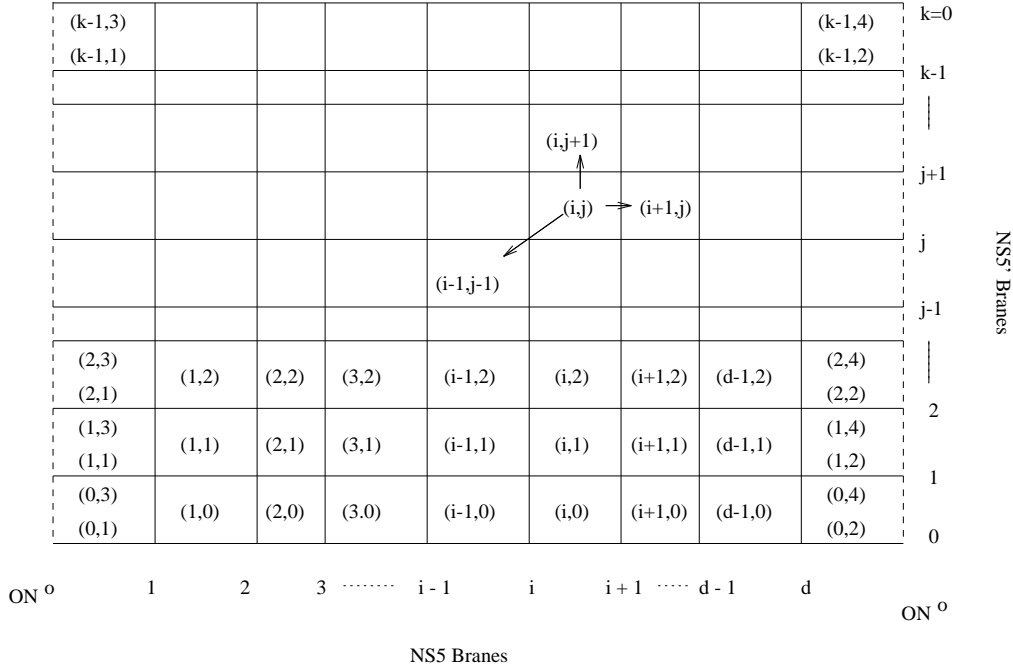,width=5.4in}}
\caption{The Brane Box Model for \G. We place $d := \frac{kk'}{\delta}$
NS5 branes in between 2 parallel ON$^0$-planes and $k$ NS5$'$ branes perpendicularly
while identifying the 0th and the $k$th circularly. Within the boxes of this grid, we
stretch D5 branes, furnishing bi-fundamental as indicated by the arrows shown.}
\label{fig:BBZD}
\end{figure}

Now we construct the brane box using the basic idea in \cite{Han-Zaf2} .
Let us focus on the case of $\delta := (k,2k')$ being even where we have
$4k$ 1-dimensional irreducible representations and $k(\frac{k'k}{(k,2k')}-1)$ 2-dimensionals.
We place 2 ON$^0$ planes vertically at two sides. Between them we place 
$\frac{kk'}{\delta}$ vertically parallel NS5 branes 
(which give the structure of $D_{\frac{kk'}{\delta}}$). 
Next we place $k$ NS5$'$ branes horizontally (which give the
structure of $Z_k$) and identify the $k$th with the zeroth. 
This gives us a grid of $k(\frac{kk'}{\delta}+1)$ boxes. Next we put $N$ D5 branes with
positive charge and $N$ with negative charge in those grids.
Under the decomposition (\ref{decomp}), we can connect the structure of singularity to
the structure of Brane Box Model by placing  the irreducible representations into the grid of boxes
\`{a} la \cite{Han-Zaf,Han-Ura} as follows (the setup is shown in
\fref{fig:BBZD}).

First we place the $4k$ 1-dimensionals at the two sides such that those boxes each 
contains two: at the left we have $(l'=0,h'=1)_1$ and $(l'=0,h'=3)_1$ 
at the lowest box and with the upper boxes containing subsequent increments on $l'$.
Therefore we have the list, in going up the boxes,
$\{ (0,1)_1~\&~(0,3)_1; 
(1,1)_1 ~\&~ (1,3)_1; (2,1)_1 ~\&~ (2,3)_1; ... (k-1,1)_1 ~\&~ (k-1,3)_1\}$. 
The right side has a similar list: 
$\{ (0,2)_1 ~\&~ (0,4)_1; 
(1,2)_1 ~\&~ (1,4)_1; (2,2)_1 ~\&~ (2,4)_1; ... (k-1,2)_1 ~\&~ (k-1,4)_1\}$. 
Into the middle grids we place the 2-dimensionals, one to a box, such that the bottom
row consists of 
$\{(d=1,l=0)_2,(2,0)_2,(3,0)_2,...(\frac{kk'}{\delta}-1,0)_2 \}$ 
from left to right. And as we go up we increment $l$ until $l=k-1$ ($l=k$ is
identified with $l=0$ due to our compactification).
Now we must check the consistency condition. We choose the bi-fundamental directions
according to the conventions in \cite{Han-Zaf,Han-Ura}, i.e., East, North and Southwest.
The consistency condition is that for the irreducible representation in box $i$, forming the tensor product with
the {\bf 3} chosen in (\ref{decomp}) should be the tensor sum of the irreducible representations
of the neighbours in the 3 chosen directions, i.e.,
\begin{equation}
{\bf 3} \otimes R_i = \bigoplus\limits_{j\in{\rm Neighbours}} R_j
\label{consistency}
\end{equation}
Of course this consistency condition is precisely (\ref{aij}) in a different
guise and
checking it amounts to seeing whether the Brane Box Model gives the
same quiver theory as does the geometry, whereby showing the equivalence 
between the two methods.
Now the elaborate tabulation in \sref{subsec:decomp} is seen to be not in vain;
let us check (\ref{consistency}) by column in the brane box as in 
\fref{fig:BBZD}.
For the $i$th entry in the leftmost column, containing $R_i=(l',1~{\rm or}~3)$,
we have $R_i \otimes {\bf 3} = (l',1~{\rm or}~3)_1 \otimes ((1,1)_1 \oplus
(1,0)_2) = (l'+1,1~{\rm or}~3)_1 \oplus (1,l')_2$. The righthand side is
precisely given by the neighbour of $i$ to the East and to the North and since
there is no Southwest neighbour, consistency (\ref{consistency}) holds for
the leftmost column. A similar situation holds for the rightmost column,
where we have ${\bf 3} \otimes (l',2~{\rm or}~4) = (l'+1,2~{\rm or}~4)_1
\oplus (\frac{kk'}{\delta}-1,l'-1)_2$, the neighbour to the North and the 
Southwest.

Now we check the second column, i.e., one between the first and second NS5-branes.
For the $i$th entry $R_i = (1,l)_2$, after tensoring with the {\bf 3},
we obtain $(1,l+1)_2  \oplus (l+1,l+0)_2 \oplus ((l+0-1,1)_1 \oplus
(l+0-1,3)_1)$, which are the irreducible representations precisely in the 3 neighbours: respectively
East, North and the two 1-dimensional in the Southwest. Whence 
(\ref{consistency}) is checked. Of course a similar situation occurs for the
second column from the right where we have ${\bf 3} \otimes 
(R_i = (\frac{kk'}{\delta}-1,l)_2) = (\frac{kk'}{\delta}-1,l+2)_2 \oplus
(\frac{kk'}{\delta}-1-1,l-1)_2 \oplus ((l,2)_1 \oplus (l,4)_1)$, or
respectively the neighbours to the North, Southwest and the East.

The final check is required of the interior box, say $R_i = (d,l)_2$.
Its tensor with {\bf 3} gives $(d,l+1)_2 \oplus (d-1,l-1)_2 \oplus
(d+1,l)_2$, precisely the neighbours to the North, Southwest and East.

\subsection{The Inverse Problem}
A natural question arises from our quest for the correspondence
between brane box constructions and branes as probes: is such a
correspondence bijective? Indeed if the two are to be related by some
T Duality or generalisations thereof, this bijection would be necessary.
Our discussions above have addressed one direction: given a \G singularity,
we have constructed a consistent Brane Box Model. Now we must ask
whether given  such a configuration with $m$ NS5 branes between two
ON$^0$  planes and $k$ NS5$'$ branes according to \fref{fig:BBZD},
could we find a unique \G orbifold which corresponds thereto?
The answer fortunately is in the affirmative and is summarised in the following:
\begin{proposition}
For $\frac{2k'}{(k,2k')}$ being even\footnote{Which is the case upon which
we focus.}, there exists a bijection\footnote{Bijection in the sense that given
a quiver theory produced from one picture there exists a unique method in the
other picture which gives the same quiver.} between the Brane Box Model and
the D3 brane-probes on the orbifold for the
group $G := Z_k \times D_{k'} \cong Z_k \rtimes D_{m:=\frac{kk'}{(k,2k')}}$.
In particular
\begin{itemize}
\item (I) Given $k$ and $k'$, whereby determining $G$ and hence the orbifold theory,
one can construct a unique Brane Box Model;
\item (II) Given $k$ and $m$ with the condition that $k$ is a divisor of $m$,
where $k$ is  the number of NS5 branes perpendicular to $ON^0$ planes and 
$m$ the number of NS5 branes between two  $ON^0$ planes
as in \fref{fig:BBZD}, one can extract a unique orbifold theory.
\end{itemize}
\end{proposition}

Now we have already shown (I) by our extensive discussion in the previous
sections. Indeed, given integers $k$ and
$k'$, we have twisted $G$ such that it is characterised by $k$ and
\begin{equation}
\label{addm}
m:=\frac{kk'}{(k,2k')},
\end{equation}
two numbers that uniquely fix the brane configuration.
The crux of the remaining direction (II) seems to be the issue whether we could,
given $k$ and $m$, ascertain the values
of $k$ and $k'$ uniquely? For if so then our Brane Box Model, which is solely
determined by $k$ and $m$, would be uniquely mapped to a \G orbifold, characterised
by $k$ and $k'$. We will show below that though this is not so and $k$ and $k'$ cannot
be uniquely solved, it is still true that $G$ remains unique. Furthermore, we will
outline the procedure by which we can find convenient choices of $k$ and $k'$ that
describe $G$.

Let us analyse this problem in more detail.
First we see that $k$, which determines the $Z_k$ in $G$, remains unchanged.
Therefore our problem is further reduced to: given $m$, is there a unique
solution of $k'$ at fixed $k$? We write $k,k',m$ as:

\begin{equation}
\label{kk}
\begin{array}{c}
k=2^{q}l f_2  \\
k'=2^{p} l f_1 \\
m=2^{n} f_3
\end{array}
\end{equation}

where with the extraction of all even factors, $l,f_1$ and $f_2$
are all odd integers and $l$ is the greatest common divisor of $k$ and $k'$ so that
$f_1,f_2$ are coprime. What we need to know are $l,f_1$ and $p$ given $k,q,n$ and $f_3$.
The first constraint is that $\frac{2k'}{(k,2k')}=\even$, a condition on which our
paper focuses. This immediately yields the inequality $p\geq q$. The definition of
$m$ (\ref{addm}) above further gives
\[
2^{n} f_3=m=2^p l f_1 f_2= 2^{p-q}k f_1.
\]
From this equation, we can solve
\begin{equation}
\label{OI1}
p=n,~~~~~~~f_1=\frac{m}{2^{p-q}k}
\end{equation}
Now it remains to determine $l$. However, the solution for
$l$ is not unique. For example, if we take $l=l_1 l_2$ and $(l_2,f_1)=1$, then
the following set $\{\tilde{k},\tilde{k'}\}$ will give same $k,m$:
\[
\begin{array}{c}
\tilde{k}=k=2^{q}l_1 l_2 f_2  \\
\tilde{k'}=2^{p} l_1 f_1 \\
m=2^{n} f_3
\end{array}
\]
This non-uniqueness in determining $k,k'$ from $k,m$ may at first seem discouraging.
However we shall see below that different pairs of $\{k,k'\}$ that give the
same $\{k,m\}$ must give the same group $G$.

We first recall that $G$ can be written as
$Z_k \rtimes D_{m=\frac{kk'}{(k,2k')}}$.
For fixed $k,m$ the two subgroups $Z_k$ and $D_m$ are same. For the whole group
$Z_k \rtimes D_{m=\frac{kk'}{(k,2k')}}$ be unique no matter which $k'$ we choose we just need to show that the algebraic relation which generate 
$Z_k \rtimes D_{m=\frac{kk'}{(k,2k')}}$ from $Z_k$ and $D_m$ is same. For that,
 we recall from the proposition
in section \sref{subsec:H}, that in twisting $G$ into its internal semi-direct form,
the crucial relation is
\[
\alpha \tilde{\gamma}=\tilde{\beta}^{\frac{2k'}{(k,2k')}} \tilde{\gamma} \alpha
\]
Indeed we observe that $\frac{k'}{(k,2k')}= \frac{m}{k}$ where the condition
that $k$ is a divisor of $m$ makes the expression having meaning. Whence given $m$ and $k$,
the presentation of $G$ as $Z_k \rtimes D_m$ is uniquely fixed, and hence $G$
is uniquely determined. This concludes our demonstration for the above proposition.

Now the question arises as to what values of $k$ and $k'$ result in the
same $G$ and how the smallest pair (or rather, the smallest $k'$ since
$k$ is fixed) may be selected. In fact our discussion
above prescribes a technique of finding such a pair. First we
solve $p,f_1$ using (\ref{OI1}), then we find the largest factor $h$ of $k$ which
satisfies $(h,f_1)=1$. The smallest value of $k'$ is then such that
$l=\frac{k}{h}$ in (\ref{kk}).
Finally, we wish to emphasize
that the bijection we have discussed is not true for arbitrary $\{m,k\}$ and we
require that $k$ be a divisor of $m$ as is needed in demonstration of the
proposition. Indeed, given $m$ and $k$ which do not satisfy
this condition, the 1-1 correspondence between the Brane Box Model and the 
orbifold singularity is still an enigma and will be 
left for future labours.

\section{Conclusions and Prospects} \label{sec:conc}
We have briefly reviewed some techniques in two contemporary 
directions in the construction
of gauge theories from branes, namely branes as geometrical probes on orbifold
singularities or as constituents of configurations of D branes stretched between
NS branes. 
Some rudiments in the orbifold procedure,
in the brane setup of
${\cal N}=2$ quiver theories of the $\widehat{D_k}$ type as well as in the
${\cal N}=1$ $Z_k \times Z_{k'}$ Brane Box Model have been introduced.
Thus inspired, we have
constructed the Brane Box Model for an infinite series of
non-Abelian finite subgroups of $SU(3)$, by combining some methodology
of the aforementioned brane setups.

In particular, we have extensively studied the properties, especially the
representation and character theory of the intransitive collineation group 
$G := Z_k \times D_{k'} \subset SU(3)$,
the next simplest group after $Z_k \times Z_{k'}$ and a natural extension thereof.
From the geometrical perspective, this amounts to the study of Gorenstein
singularities of the type $\C^3 / G$ with the $Z_k$ acting on the first two
complex coordinates of $\C^3$ and $D_{k'}$, the last two.

We have shown why na\"{\i}ve Brane Box constructions for $G$ fail (and indeed
why non-Abelian groups in general may present difficulties). It is only after
a ``twist'' of $G$ into a semi-direct product form $Z_k \rtimes D_{\frac{kk'}{(k,2k')}}$,
an issue which only arises because of the non-Abelian nature of $G$, that
the problem may be attacked. For $\frac{2k'}{(k,2k')}$ even, we have successfully
established a consistent Brane Box Model. The resulting gauge theory is that of
$k$ copies of $\widehat{D}$-type quivers circularly arranged (see \fref{fig:BBZD}).
However for $\frac{2k'}{(k,2k')}$ odd, a degeneracy occurs and we seem to arrive at
ordinary (non-Affine) $D$ quivers, a phenomenon hinted at by some previous 
works \cite{Han-Zaf2,Han-He} but still remains elusive. Furthermore, we have
discussed the inverse problem, i.e., whether given a configuration of the
Brane Box Model we could find the corresponding branes as probes on orbifolds. 
We have shown that when $k$ is a divisor of $m$
the two perspectives are bijectively related
and thus the inverse problem can be solved.
For general $\{m,k\}$, the answer of the inverse problem is still not clear.

Many interesting problems arise and are open. Apart from clarifying the physical
meaning of ``twisting'' and hence perhaps treat the $\frac{2k'}{(k,2k')}$ odd case,
we can try to construct Brane Boxes for more generic non-Abelian groups.
Moreover, marginal couplings and duality groups thereupon may be extracted
and interpreted as brane motions; this is of particular interest because
toric methods from geometry so far have been restricted to Abelian singularities.
Also, recently proposed brane diamond models \cite{Aganagic} may be combined with
our techniques to shed new insight. Furthermore during the preparation of this
manuscript, a recent paper that deals with brane configurations for $\C^3/\Gamma$
singularities for non-Abelian $\Gamma$ (i.e the $\Delta$ series in $SU(3)$)
by $(p,q)$5-brane webs has come to our attention \cite{Muto2}.
We hope that our construction, as the  Brane Box Model realisation of a non-Abelian
orbifold theory in dimension 3, may lead to insight in these various directions.

\section*{Acknowledgements}
{\it Catharinae Sanctae Alexandriae et Ad Majorem Dei Gloriam...\\}
We would like to extend our sincere gratitude to A. Kapustin, 
A. Karch, A. Uranga
and A. Zaffaroni for fruitful discussions. Furthermore we
would like to thank O. DeWolfe, L. Dyson, J. Erlich, A. Naqvi,
M. Serna and J. S. Song for their suggestions and help. YHH is also obliged
to N. Patten and S. Mcdougall for charming diversions.

\section*{Appendix}
Using the notation introduced in \sref{sec:group}, we see that the 
conjugation within $G$ gives
\begin{equation}
(q,\tilde{q},\tilde{n},k)^{-1}(m,\tilde{m},n,p) (q,\tilde{q},\tilde{n},k)=
\left\{  
\begin{array}{l}
  (\tilde{m}+q-\tilde{q},m-q+\tilde{q},n,2k-p) $ for $ n=0,\tilde{n}=0 \\
  (m-q+\tilde{q}, \tilde{m}+q-\tilde{q},n,2k+p) $ for $ n=0,\tilde{n}=1 \\
 (\tilde{m},m,n,-p) $ for $ n=1,\tilde{n}=0  \\
  (m,\tilde{m},n,p) $ for $ n=1,\tilde{n}=1.
        \end{array}
\right.
\label{conj}
\end{equation}
Also, we present the multiplication rules in $G$ for reference:
\begin{eqnarray}
(m,\tilde{m},0,p_1)(n,\tilde{n},0,p_2) & = & (m+\tilde{n},\tilde{m}+n,1,p_2-p_1)
\nonumber \\
(m,\tilde{m},0,p_1)(n,\tilde{n},1,p_2) & = & (m+\tilde{n},\tilde{m}+n,0,p_2+p_1-k')
\nonumber \\
(m,\tilde{m},1,p_1)(n,\tilde{n},0,p_2) & = & (m+n,\tilde{m}+\tilde{n},0,p_2-p_1-k')
\nonumber \\
(m,\tilde{m},1,p_1)(n,\tilde{n},1,p_2) & = & (m+n,\tilde{m}+\tilde{n},1,p_2+p_1-k')
\label{multi}
\end{eqnarray}

First we focus on the conjugacy class of elements such that $n=0$.
From (\ref{orbit}) and (\ref{conj}), we see that if two 
elements are within the same conjugacy class, then they must have the same 
$m+\tilde{m} \bmod k$.
Now we need to distinguish between two cases:
\begin{itemize}
\item (I) if $\frac{2k'}{(k,2k')}=\even$, the orbit conditions conserve 
the parity of $p$, making even and odd $p$ belong to different conjugacy classes;
\item (II) if $\frac{2k'}{(k,2k')}=\odd$, the orbit conditions change $p$ and 
we find that all $p$ belong to the same conjugacy class 
they have the same value for $m+\tilde{m}$.
\end{itemize}
In summary then, for $\frac{2k'}{(k,2k')}=\even$, we have $2k$ 
conjugacy classes each of which has $\frac{k'k}{(k,2k')}$ elements;
for $\frac{2k'}{(k,2k')}=\odd$, we have $k$ 
conjugacy classes each of which has $\frac{2k'k}{(k,2k')}$ elements. 

Next we analyse the conjugacy class corresponding to $n=1$.
For simplicity, we divide the interval $[0,k)$ by factor $(k,2k')$ and define 
\[
V_{i} = \left[\frac{ik}{(k,2k')},\frac{(i+1)k}{(k,2k')}\right)
\]
with $i=0,...,(k,2k')-1$. 
Now from (\ref{orbit}), we can always fix $m$ to belong $V_{0}$. 
Thereafter, $\tilde{m}$ and $p$ can change freely within $[0,k)/[0,2k')$. 
Again, we have two different cases.
(I) If $\frac{2k'}{(k,2k')}=\even$, for every subinterval $V_{i}$ 
we have $2k_{0}$ (we define $k_0:=2\frac{k}{(k,2k')}$) conjugacy classes each containing
only one element, namely, 
\[
(m,\tilde{m}=m+\frac{ik}{(k,2k')},n=1,p=k'-\frac{ik'}{(k,2k')} {\rm ~or~}
2k'-\frac{ik'}{(k,2k')}).
\]
Also we have a total of 
$k_{0}\frac{2k'-2}{2}+\frac{k_{0}(k_{0}-1)}{2}
2k'=k_{0}(k'-1)+k'k_{0}^2+k_{0}k'=k'k_{0}^2-k_{0}$
conjugacy classes of 2 elements, namely
$(m,\tilde{m},n=1,p)$ and
$(\tilde{m}-\frac{ik}{(k,2k')},m+\frac{ik}{(k,2k')},n=1,-p-i\frac{2k'}{(k,2k')})$.
Indeed, the total number of conjugacy classes is 
$2k+(k,2k')(2k_{0})+(k,2k')(k'k_{0}^2-k_{0})=4k+k(\frac{k'k}{(k,2k')}-1)$, giving
the order of $G$ as expected.
Furthermore, there are $4k$ 1-dimensional irreducible representations 
and $k(\frac{k'k}{(k,2k')}-1)$ 2-dimensional irreducible representations. This is consistent since
$\sum_i\dim{\bf r}_i = 1^2\cdot 4k+2^2\cdot k(\frac{k'k}{(k,2k')}-1)=
\frac{4k'k^2}{(k,2k')} = |G|$.

We summarize case (I) into the following table:
\[
\begin{array}{c|c|c|c}
      	& C_{n=0}^{m+\tilde{m}(\bmod k),p=\odd / \even}  &
        C_{n=1,V_{i}}^{\tilde{m}=m+\frac{ik}{(k,2k')},p=
			(k'-\frac{ik'}{(k,2k')}) /
        		(2k'-\frac{ik'}{(k,2k')})} 
	& C_{n=1,V_{i}}^{(m,\tilde{m},p)=
		(\tilde{m}-\frac{ik}{(k,2k')},
		m+\frac{ik}{(k,2k')},-p-i\frac{2k'}{(k,2k')})} \\ \hline
|C| & \frac{k'k}{(k,2k')}  & 1 & 2 \\ \hline
\#C &  2k  & 2k & k(\frac{k'k}{(k,2k')}-1)
\end{array}
\]

Now let us treat case (II), where $\frac{2k'}{(k,2k')}$ is odd (note that
in this case we must have $k$ even). Here, for $V_{i}$ and $i$ even, the situation
is as (I) but for $i$ odd there are no one-element conjugacy classes. We tabulate the
conjugacy classes in the following:
\[
\begin{array}{c|c|c|c}
	& C_{n=0}^{m+\tilde{m}(\bmod k),\rm{any~}p}  &
        C_{n=1,V_{i},i=\even}^{\tilde{m}=m+\frac{ik}{(k,2k')},
		p=(k'-\frac{ik'}{(k,2k')})
        	/(2k'-\frac{ik'}{(k,2k')})}
	& C_{n=1,V_{i}}^{(m,\tilde{m},p)=
		(\tilde{m}-\frac{ik}{(k,2k')},
		m+\frac{ik}{(k,2k')},-p-i\frac{2k'}{(k,2k')})} \\ \hline
|C| & \frac{2k'k}{(k,2k')}  & 1 & 2 \\ \hline
\#C &  k  & 2\frac{k}{2}=k & \frac{(k,2k')}{2}[(k'k_{0}^2-k_{0})+k'k_{0}^2]=
k(\frac{k'k}{(k,2k')}-\frac{1}{2})
\end{array}
\]


\begin{thebibliography}{9}
\bibitem{Han-Wit} A. Hanany and E. Witten, ``Type IIB Superstrings, BPS monopoles, and 
        Three-Dimensional Gauge Dynamics,'' hep-th/9611230.

\bibitem{Giveon} A. Giveon and D. Kutasov, ``Brane Dynamics and Gauge Theory,''
        hep-th/9802067.

\bibitem{Boer} Jan de Boer, Kentaro Hori, Hirosi Ooguri, Yaron Oz and Zheng Yin,
        ``Mirror Symmetry in Three-dimensional Gauge Theories, $SL(2,Z)$ and
        D-Brane Moduli Spaces,'' hep-th/9612131.

\bibitem{Kapustin} A. Kapustin, ``$D_n$ Quivers from Branes,'' hep-th/9806238.

\bibitem{P-Zaf}  M. Porrati, A. Zaffaroni, ``M-Theory Origin of Mirror Symmetry	 in Three Dimensional Gauge Theories'', Nucl.Phys. B490 (1997) 107-120,
	hep-th/9611201.

\bibitem{IS}  K. Intriligator, N. Seiberg, ``Mirror Symmetry in Three 
	Dimensional Gauge Theories,'' Phys.Lett. B387 (1996) 513-519,
	hep-th/9607207.

\bibitem{Elitzur}  S. Elitzur, A. Giveon and D. Kutasov, ``Branes and N=1 
	Duality in String Theory,'' Phys.Lett. B400 (1997) 269-274, hep-th/9702014.  \\
	S. Elitzur, A. Giveon, D. Kutasov, E. Rabinovici and A. Schwimmer,
        ``Brane dynamics and $N=1$ supersymmetric gauge theory,'' Nucl.Phys.B 
        {\bf 505}(1997) 202-250.

\bibitem{Mlift} E. Witten, ``Solutions of Four-Dimensional Field Theories Via M Theory,''
        hep-th/9703166.

\bibitem{Mirror} S. Katz, P. Mayr and C. Vafa, ``Mirror symmetry and Exact Solution 
        of 4D N=2 Gauge Theories I,'' hep-th/9706110.

\bibitem{Quiver} M. Douglas and G. Moore, ``D-Branes, Quivers, and ALE Instantons,''
        hep-th/9603167.

\bibitem{Karch} A. Karch, ``Field Theory Dynamics from Branes in String Theory,''
        hep-th/9812072.  \\
        Andreas Karch, Dieter Lust, Douglas J. Smith,
        ``Equivalence of Geometric Engineering 
	and Hanany-Witten via Fractional Branes'', hep-th/9803232.

\bibitem{Karl}  Karl Landsteiner, Esperanza Lopez, David A. Lowe, `` N=2 Supersymmetric 
	Gauge Theories, Branes and Orientifolds'',
	Nucl.Phys. B507 (1997) 197-226. hep-th/9705199. \\
	 Karl Landsteiner, Esperanza Lopez, ``New Curves from Branes'',
	Nucl.Phys. B516 (1998) 273-296, hep-th/9708118.

\bibitem{B-Karch} Ilka Brunner, Andreas Karch, ``Branes at Orbifolds 
	versus Hanany Witten in Six Dimensions'', JHEP 9803 (1998) 003, hep-th/9712143.

\bibitem{Park} J. Park, A. M. Uranga, `` A Note on Superconformal 
	N=2 theories and Orientifolds'', Nucl.Phys. B542 (1999) 139-156, hep-th/9808161.


\bibitem{Erlich} J. Erlich, A. Hanany, and A. Naqvi, ``Marginal Deformations 
        from Branes,'' hep-th/9902118.

\bibitem{Conf} S. Kachru and E. Silverstein, ``4D Conformal Field Theories
        and Strings on Orbifolds,'' hep-th/9802183.\\
        A. Lawrence, N. Nekrasov and C. Vafa, ``On Conformal Field
        Theories in Four Dimensions,'' hep-th/9803015. \\
	Michael Bershadsky, Zurab Kakushadze, Cumrun Vafa,
	``String Expansion as Large N Expansion of Gauge Theories'',
	Nucl.Phys. B523 (1998) 59-72, hep-th/9803076.

\bibitem{Quiver1}  M. Douglas and B. Greene, ``Metrics on D-brane Orbifolds,''
        hep-th/9707214.\\
        M. Douglas, B. Greene, and D. Morrison, 
        ``Orbifold Resolution by D-Branes,'' hep-th/9704151.

\bibitem{Han-He} A. Hanany and Y.-H. He, ``Non-Abelian Finite Gauge Theories,''
        hep-th/9811183.

\bibitem{Muto1}	T. Muto, ``D-branes on Three-dimensional Nonabelian Orbifolds'', J.High
	Energy Phys. {\bf 02}(1999)008, hep-th/9811258.

\bibitem{Han-Zaf} A. Hanany and A. Zaffaroni, ``On the Realisation of
	Chiral Four-Dimenaional Gauge Theories Using Branes,'' hep-th/9801134.

\bibitem{Han-Ura} A. Hanany and A. Uranga, ``Brane Boxes and Branes on Singularities,''
        hep-th/9805139.

\bibitem{Han-S} Amihay Hanany, Matthew J. Strassler and Angel M. Uranga, 
        ``Finite Theories and Marginal Operators on the Brane,''
        hep-th/9803086,JHEP 9806 (1998) 011.

\bibitem{Leigh} R. Leigh and M. Rozali, ``Brane Boxes, Anomalies, Bending and
        Tadpoles,'' hep-th/9807082.

\bibitem{Kutasov}David Kutasov, ``Orbifolds and Solitons'',
 	Phys.Lett. B383(1996) 48-53, hep-th/9512145. 

\bibitem{Sen} A. Sen, ``Duality and Orbifolds,'' hep-th/9604070. \\
        A. Sen, ``Stable Non-BPS Bound States of BPS D-branes'', hep-th/9805019.

\bibitem{Han-Zaf2}  A. Hanany and A. Zaffaroni, ``Issues on Orientifolds: On the Brane
        Construction of Gauge Theories with $SO(2n)$ Global Symmetry,'' hep-th/9903242.

\bibitem{Dog1} Duiliu-Emanuel Diaconescu, Michael R. Douglas, Jaume Gomis,
	``Fractional Branes and Wrapped Branes'', JHEP 9802 (1998) 013,
	hep-th/9712230.

\bibitem{Karch1} Andreas Karch, Dieter Lust, Douglas J. Smith,
	``Equivalence of Geometric Engineering and Hanany-Witten
	via Fractional Branes'', Nucl.Phys. B533 (1998) 348-372, hep-th/9803232.

\bibitem{Greene}Brian R. Greene, C. I. Lazaroiu, Mark Raugas, ``D-branes on 
	Nonabelian Threefold Quotient Singularities'', hep-th/9811201.

\bibitem{Su4} A. Hanany and Y.-H. He, ``A Monograph on the 
        Classification of the Discrete Subgroups of $SU(4)$,'' hep-th/9905212.

\bibitem{He-Song} Y.-H. He and J. S. Song, ``Of McKay Correspondence,
        Non-linear Sigma-model, and Conformal Field Theory,''
        hep-th/9903056.

\bibitem{LRA}L. E. Ibanez, R. Rabadan, A. M. Uranga, ``Anomalous U(1)'s in Type I and 
	Type IIB D=4, N=1 string vacua'', Nucl.Phys. B542 (1999) 112-138. 
	hep-th/9808139.

\bibitem{Alperin} J. Alperin and R. Bell, ``Groups and Representation,''
        GTM162, Springer-Verlag NY (1995).

\bibitem{Aganagic} M. Aganagic, A. Karch, D. Lust and A. Miemiec
	``Mirror Symmetries for Brane Configurations and Branes at
          Singularities,'' hep-th/9903093.

\bibitem{Muto2} T. Muto, ``Brane Configurations for Three-dimensional 
	Nonabelian Orbifolds,'' hep-th/9905230.

\end{thebibliography}
\end{document}